\documentclass{aa}

\usepackage{graphicx}
\usepackage{txfonts}
\begin{document}

   \title{Asteroseismology of the $\delta$ Scuti star HD 50844}

   \author{X. H. Chen\inst{1,2,3}, Y. Li\inst{1,2}, X. J. Lai\inst{1,2}, and T. Wu\inst{1,2}}

   \institute{Yunnan Observatories, Chinese Academy of Sciences, P.O. Box 110, Kunming 650011, China\\
             \email{chenxinghao@ynao.ac.cn; ly@ynao.ac.cn}
         \and
             Key Laboratory for the Structure and Evolution of Celestial Objects, Chinese Academy of Sciences, P.O. Box 110, Kunming 650011, China
         \and
             University of Chinese Academy of Sciences, Beijing 100049, China\\
             }

   \date{January 28, 2016}


  \abstract
   {}
   {We aim to probe the internal structure and investigate more detailed information of the $\delta$ Scuti star HD 50844 with asteroseismology.}
   {We analyse the observed frequencies of the $\delta$ Scuti star HD 50844 obtained by Balona (2014), and search for possible multiplets based on the rotational splitting law of g-mode. We tried to disentangle the frequency spectra of HD 50844 by means of the rotational splitting only. We then compare them with theoretical pulsation modes, which correspond to stellar evolutionary models with various sets of initial metallicity and stellar mass, to find the best-fitting model.}
   {There are three multiplets including two complete triplets and one incomplete quintuplet, in which mode identifications for spherical harmonic degree $l$ and azimuthal number $m$ are unique. The corresponding rotational period of HD 50844 is found to be 2.44$^{+0.13}_{-0.08}$ days. The physical parameters of HD 50844 are well limited in a small region by three modes identified as nonradial ones ($f_{11}$, $f_{22}$, and $f_{29}$) and by the fundamental radial mode ($f_{4}$). Our results show that the three nonradial modes ($f_{11}$, $f_{22}$, and $f_{29}$) are all mixed modes, which mainly represent the property of the helium core. The fundamental radial mode ($f_{4}$) mainly represents the property of the stellar envelope. In order to fit these four pulsation modes, both the helium core and the stellar envelope must be matched to the actual structure of HD 50844. Finally, the mass of the helium core of HD 50844 is estimated to be 0.173 $\pm$ 0.004 $M_{\odot}$ for the first time. The physical parameters of HD 50844 are determined to be $M=$ 1.81 $\pm$ 0.01 $M_{\odot}$, $Z=$ 0.008 $\pm$ 0.001. $T_{\rm eff}=$ 7508 $\pm$ 125 K, log$g=$ 3.658 $\pm$ 0.004, $R=$ 3.300 $\pm$ 0.023 $R_{\odot}$, $L=$ 30.98 $\pm$ 2.39 $L_{\odot}$.}
   {}

   \keywords{asteroseismology -
          stars: individual: HD 50844 -
          $\delta$ Scuti star
               }
   \maketitle
%
\section{Introduction}
The $\delta$ Scuti stars are a class of pulsating stars falling in the HR diagram where the main sequence overlaps the lower extension of the Cepheid instability strip (Breger 2000; Aerts et al. 2010). They are in the core hydrogen burning or shell hydrogen burning stage (Aerts et al. 2010), with masses from 1.5 $M_{\odot}$ to 2.5 $M_{\odot}$ (Aerts et al. 2010) and pulsation periods from 0.5 to 6 hours (Breger 2000).  Their pulsations are driven by the $\kappa$ mechanism (Baker $\&$ Kippenhahn 1962, 1965; Zhevakin 1963; Li $\&$ Stix 1994) in the second partial ionization zone of helium. Some of $\delta$ Scuti stars show multi-period pulsations, such as 4 Cvn (Breger et al. 1999), FG Vir (Breger et al. 2005), and HD 50870 (Mantegazza et al. 2012), and are therefore good candidates for asteroseismological studies.

HD 50844 was discovered to be a $\delta$ Scuti variable star by Poretti et al. (2005) during their preparatory work for the CoRoT mission. The basic physical parameters of HD 50844 were also obtained by Poretti et al. (2005) from Str$\ddot{o}$mgren photometry. They are listed as follows: $T_{\rm eff}=$ 7500 $\pm$ 200K, log$g$=3.6 $\pm$ 0.2, and [Fe/H] = $-0.4$ $\pm$ 0.2. High-resolution spectroscopic observations obtained with the FEROS instrument mounted on the 2.2-m ESO/MPI telescope at La Silla resulted in the value of $\upsilon\sin i$ = 58 $\pm$ 2 km $\rm s^{-1}$ and the inclination angle $i=$ 82 $\pm$ 4 deg (Poretti et al. 2009).

HD 50844 was observed from 2 February 2007 to 31 March 2007 ($\Delta t $= 57.61 d) by CoRoT during the initial run (IR01).  Detailed frequency analysis of the observed timeseries by Poretti et al. (2009) revealed very dense frequency signals in the range of 0-30 ${\rm d}^{-1}$. In particular, they identified the frequency 6.92 ${\rm d}^{-1}$ with the largest amplitude as the fundamental radial mode by combining spectroscopic and photometric data. Meanwhile, very high-degree oscillation modes (up to $l=14$) were identified by Poretti et al. (2009) with the software FAMIAS (Zima 2008) to fit the line profile variations (Mantegazza 2000). Based on an independent analysis, Balona (2014) arrived a conclusion that a normal mode density might be existent for the CoRoT timeseries of HD 50844. He extracted a total of 59 significant oscillation modes from the CoRoT timeseries.

Asteroseismology is a powerful tool to investigate the internal structure of pulsating stars that show rich pulsation modes in observations, such as 44 Tau (Civelek et al. 2001; K{\i}rb{\i}y{\i}k et al. 2003; Garrido et al. 2007; Lenz et al. 2008, 2010) and $\alpha$ Oph (Zhao et al. 2009; Deupree et al. 2012). Mode identifications are very important for asteroseismic studies of $\delta$ Scuti stars. Any eigenmode of stellar nonradial oscillations can be characterized by its radial order $k$, spherical harmonic degree $l$, and azimuthal number $m$ (Christensen-Dalsgaard 2003). For $\delta$ Scuti stars, there are usually only a few observed modes to be identified, such as FG Vir (Daszy$\acute{\rm n}$ska-Daszkiewicz et al.2005; Zima et al. 2006) and 4 Cvn (Castanheira et al. 2008; Schmid et al. 2014). Observed frequencies of a pulsating star could be compared with results of theoretical models only if their values ($l$, $m$) have been determined in advance.

For a rotating star, a non-radial oscillation mode will split into $2l+1$ different components. According to the asymptotic theory of stellar oscillations, the $2l+1$ components of one mode of ($k$, $l$) are separated by almost the same spacing for a slowly rotating star. In our work, we try to identify the observed frequencies obtained by Balona (2014) based on the rotational splitting law of g-mode. Then we compute a grid of theoretical models to examine whether the computed stellar models can provide a reasonable fit to the observed frequencies. In Section 2, we analyse the rotational splitting of the observational data. In Section 3, we describe our stellar models, including input physics in Section 3.1 and details of the grid of stellar models in Section 3.2. Our fitting results are analysed in Section 4. Finally, we summarize our results in Section 5.

\section{Analysis of rotational splitting}
As already pointed out by Poretti et al. (2005), HD 50844 is in the post-main sequence evolution stage with a contracting helium core and an expanding envelope. The steep gradient of chemical abundance in the hydrogen burning shell will result in a large Brunt-V$\ddot{a}$is$\ddot{a}$l$\ddot{a}$ frequency $N$ there. As a result, there are two propagation zones inside the star: one for g modes in the helium core and the other for p modes in the stellar envelope. As already pointed out by Poretti et al. (2009), most of the observed pulsation modes for HD 50844 should be gravity and mixed modes. The mixed modes are dominated by two characteristics. They have pronounced g-mode character in the helium core and p-mode character in the stellar envelope. In our work, we pay more attention to those modes having frequencies near or higher than that of the fundamental radial mode ($\nu > 75\mu$Hz). We list 40 frequencies obtained by Balona (2014) in Table 1. Errors of the observed frequencies obtained by Balona (2014) are too small, i.e., less than 0.0015 $\mu$Hz. They are not listed in Table 1.

The approximate expression for rotational splitting($\delta \nu_{k,l}$) and rotational period($P_{rot}$) for g-mode was derived by Dziembowski $\&$ Goode (1992) as
\begin{equation}
\nu_{k,l,m}-\nu_{k,l,0}=m\delta\nu_{k,l}=\frac{m}{P_{\rm rot}}(1-\frac{1}{L^{2}})+\frac{m^{2}}{P_{\rm rot}^{2}\nu_{k,l,0}}\frac{4L^{2}(2L^{2}-3)-9}{2L^{4}(4L^{2}-3)}.
\end{equation}
In Eq. (1), $L^{2}=l(l+1)$, and m ranges form $-l$ to $l$, resulting in $2l+1$ different values. Considering $\upsilon\sin i$ = 58 $\pm$ 2 km $\rm s^{-1}$ of HD 50844 (Poretti et al. 2009), the second term on the right-hand side of Eq.(1) is very small (e.g., less than 1.6$\%$ of the value of the first term for $\frac{1}{P_{\rm rot}}$ = 5 $\mu$Hz and $\nu_{k,l,0}$ = 100 $\mu$Hz). Therefore, only the first-order effect is considered in our work.

According to Eq.(1), modes with $l=1$ constitute a triplet. Modes with $l=2$ constitute a quintuplet, and modes with $l=3$ constitute a septuplet. Meanwhile, the rotational splitting of $l=1$ modes and those of $l=2$ modes and $l=3$ modes are in some certain proportion, i.e., $\delta\nu_{k,l=1} : \delta\nu_{k,l=2}$ : $\delta\nu_{k,l=3}$ = $\frac{3}{5}$ : 1: $\frac{11}{10}$ (Winget et al. 1991). Furthermore, the value of rotational splitting are very close for modes with $l\ge3$, e.g., the differences of rotational splittings between modes with $l=4$ and those with $l=3$ are about $\frac{0.033}{P_{\rm rot}}$. If one complete nontuplet is identified, these modes can be identified as modes with $l=4$. But beyond that, it is difficult to distinguish multiplets of $l=4$ and higher values from those of $l=3$. High spherical harmonic degree modes are detected in spectroscopy of several $\delta$ Scuti stars, such as HD 101158 (Mantegazza 1997) and BV Cir (Mantegazza et al. 2001). There is no complete nontuplet to be identified for HD 50844, thus modes with $l\ge4$ are not considered in our work.

Based on the above considerations, there are two properties of the frequency splitting due to rotation. Firstly, the $2l+1$ splitting frequencies of one mode are separated by nearly equal split. Secondly, rotational splittings derived from modes with different spherical harmonic degree $l$ are in specific proportion. Frequency differences ranging from 1 $\mu$Hz to 20 $\mu$Hz are searched for the observed frequencies. Possible multiplets due to rotational splitting are listed in Table 2.

It can be found in Table 2 that ($f_{21}$, $f_{22}$, $f_{23}$) and ($f_{27}$, $f_{29}$, $f_{33}$) constitute two multiplets with an averaged frequency difference $\delta\nu_{1}$ of 2.434 $\mu$Hz, and ($f_{9}$, $f_{11}$, $f_{14}$) constitute another multiplet with an averaged frequency difference $\delta\nu_{2}$ of 8.017 $\mu$Hz. It is worth to note that the ratio of $\delta\nu_{1}$ and $\delta\nu_{2}$/2 is 0.607, which agrees well with the property of g-mode rotational splitting (Winget et al. 1991). Therefore, ($f_{21}$, $f_{22}$, $f_{23}$) and ($f_{27}$, $f_{29}$, $f_{33}$) are identified as two complete triplets, which are denoted with Multiplet 1 and Multiplet 2 in Table 2. Poretti et al. (2009) performed mode identifications with the FAMIAS method. They identified  $f_{21}$ as ($l=3$, $m=3$), $f_{23}$ as ($l=3$, $m=2$), and $f_{33}$ as ($l=3$, $m=1$). Besides, ($f_{9}$, $f_{11}$, $f_{14}$) is identified as modes with $l=2$ on basis of the ratio of $\delta\nu_{1}$ and $\delta\nu_{2}$/2. Values of azimuthal number $m$ for $f_{9}$, $f_{11}$, and $f_{14}$ are then identified to be $m=(-2, 0, +2)$. Poretti et al. (2009) identified $f_{9}$ as a low $l$ mode. Moveover, the value of $\delta\nu_{k,l=1}$ is estimated to be 2.434 $\mu$Hz, $\delta\nu_{k,l=2}$ to be 4.009 $\mu$Hz and $\delta\nu_{k,l=3}$ to be 4.462 $\mu$Hz.

Frequencies of $f_{24}$ and $f_{25}$ may constitute Multiplet 3 with a frequency difference of 2.431 $\mu$Hz, which is approximate to $\delta\nu_{k,l=1}$. We may identify their spherical harmonic degree as $l=1$. When identifying their azimuthal number $m$, there are two possibilities, i.e., corresponding to modes of $m=(-1,0)$ or $m=(0,+1)$. Poretti et al. (2009) identified $f_{24}$ as ($l=5$, $m=3$).

Frequencies of $f_{1}$ and $f_{5}$ may constitute Multiplet 5 with a frequency difference of 8.072 $\mu$Hz, which is about twice of $\delta\nu_{k,l=2}$. We may identify their spherical harmonic degree as $l=2$. When identifying their azimuthal number $m$, there are three possibilities, i.e., corresponding to modes of $m=(-2,0)$, $m=(0,+2)$, or $m=(-1,+1)$.

Besides, the frequency difference between $f_{15}$ and $f_{18}$ is 3.939 $\mu$Hz and the frequency difference between $f_{35}$ and $f_{36}$ is 3.983 $\mu$Hz. Both of them are approximate to $\delta\nu_{k,l=2}$. We may identify their spherical harmonic degree as $l=2$. There are four possible identifications for their azimuthal number $m$, i.e., corresponding to modes of $m=(-2,-1)$, $m=(-1,0)$, $m=(0,+1)$, or $m=(+1,+2)$. Poretti et al. (2009) identified $f_{15}$ as ($l=8$, $m=5$), and $f_{35}$ as ($l=12$, $m=10$).

It can be noticed in Table 2 that ($f_{2}$, $f_{7}$), ($f_{12}$, $f_{20}$), and ($f_{38}$, $f_{39}$) may constitute 3 independent multiplets. The frequency difference between $f_{2}$ and $f_{7}$ is 12.228 $\mu$Hz. The frequency difference between $f_{12}$ and $f_{20}$ is 11.825 $\mu$Hz, and the frequency difference between $f_{38}$ and $f_{39}$ is 11.890 $\mu$Hz. All of these frequency differences are about three times of $\delta\nu_{k,l=2}$. We may identify their spherical harmonic degree as $l=2$. When identifying their azimuthal number $m$, there are two possibilities, i.e., corresponding to modes of $m=(-2,+1)$ or $m=(-1,+2)$. Poretti et al. (2009) identified $f_{12}$ as ($l=3$, $m=1$), and $f_{39}$ as ($l=14$, $m=12$).

Three frequencies of $f_{3}$, $f_{6}$ and $f_{8}$ may constitute Multiplet 11. The frequency difference 8.985 $\mu$Hz between $f_{3}$ and $f_{6}$ is about twice of $\delta\nu_{k,l=3}$, and the frequency difference 13.412 $\mu$Hz between $f_{6}$ and $f_{8}$ is about three times of $\delta\nu_{k,l=3}$. Their spherical harmonic degree $l$ can be determined to be $l=3$. There are two possible identifications for their azimuthal number $m$, i.e., corresponding to modes of $m=(-3,-1,+2)$ or $m=(-2,0,+3)$.

Another three frequencies of $f_{10}$, $f_{17}$ and $f_{19}$ may constitute Multiplet 12. The frequency difference 17.253 $\mu$Hz between $f_{10}$ and $f_{17}$ is about four times of $\delta\nu_{k,l=3}$, and the frequency difference 4.302 $\mu$Hz between $f_{17}$ and $f_{19}$ is approximate to $\delta\nu_{k,l=3}$. We may identify their spherical harmonic degree as $l=3$. When identifying their azimuthal number $m$, there are two possibilities, i.e., corresponding to modes of $m=(-3,+1,+2)$ or $m=(-2,+2,+3)$. Poretti et al. (2009) identified $f_{10}$ as ($l=5$, $m=0$), and $f_{17}$ as ($l=11$, $m=7$).

Four frequencies of $f_{26}$, $f_{28}$, $f_{34}$ and $f_{37}$ may constitute Multiplet 13. It can be noticed in Table 2 that $f_{26}$, $f_{28}$ and $f_{34}$ have a frequency difference of about $\delta\nu_{k,l=3}$ in either pair. The frequency difference 13.331 $\mu$Hz between $f_{34}$ and $f_{37}$ is about three times of $\delta\nu_{k,l=3}$. We may identify their spherical harmonic degree as $l=3$. There are two possible identifications for their azimuthal number $m$, i.e., corresponding to modes of $m=(-3,-2,-1,+2)$ or $m=(-2,-1,0,+3)$.

It can be noticed in Table 2 that there are slight differences for the rotational splittings in different multiplets. This may be due to the deviations from the asymptotic formula (e.g., Multiplet 5 and Mutiplet 6). There are also slight differences in the same multiplet (e.g., in Multiplet 2). The present frequency resolution 0.2 $\mu$Hz may be the main reason for this difference. Besides, there are only two components in Multiplet 3, 5, 6, 7, 8, 9 and 10. Different physical origins like the large separation led by the so-called island modes (Garc{\'{\i}}a Hern{\'a}ndez et al. 2013; Ligni{\`e}res et al. 2006) are also possible due to the assumptions we adopted in our approach.

There are six unidentified frequencies ($f_{13}$, $f_{16}$, $f_{30}$, $f_{31}$, $f_{32}$, and $f_{40}$) for absence of frequency splitting. It can be noticed in Table 1 that $f_{40}$ is far from other observed frequencies, so that it could not be identified on basis of rotational splitting law. Besides, frequency difference between $f_{13}$ and $f_{15}$ is 4.460 $\mu$Hz, which agrees with $\delta\nu_{k,l=3}$. It can be noticed in Table 2 that frequency difference between $f_{15}$ and $f_{18}$ is 3.939 $\mu$Hz, which agrees with $\delta\nu_{k,l=2}$. There are two possible identifications for $f_{15}$, i.e., as a mode with $l=3$ or $l=2$. There are six possibilities for the former case, i.e., corresponding to modes of $m=(-3,-2)$, $(-2,-1)$, $(-1,0)$, $(0,+1)$, $(+1,+2)$, or $(+2,+3)$. There are four possibilities for the later case, just as listed in Table 2. The frequency difference between $f_{16}$ and $f_{21}$ is 13.184 $\mu$Hz, which is about three times of $\nu_{k,l=3}$. It can be found in Table 2 that $f_{21}$ is identified as one component of a complete triplet (Multiplet 1). In Multiplet 1, the frequency difference between $f_{21}$ and $f_{22}$ agrees well with the difference between $f_{22}$ and $f_{23}$. Besides, those large differences in amplitude for $f_{21}$, $f_{22}$, and $f_{23}$ agree well with the inclination angle $i=$ 82 $\pm$ 4 deg (Poretti et al. 2009) according to the relation derived by Gizon $\&$ Solanki (2003). Furthermore, spherical harmonic degree $l$ represents the number of nodal lines on the spherical surface. For higher spherical harmonic degree $l$, the sphere will be divided into more zones. Due to the geometrical cancellation, modes with low degree $l$ are easier observed. Four frequencies of $f_{30}$, $f_{31}$,$f_{32}$, and $f_{33}$ are too much close to each other. It can be noticed in Table 2 that $f_{33}$ has already been identified as one component of a triplet (Multiplet 2). Frequency spacing between theoretical pulsation modes with $l=1$ is very large, thus $f_{30}$, $f_{31}$, and $f_{32}$ could not be identified as modes with $l=1$. Besides, only one component is observed. It is difficult to identify their spherical harmonic degree $l$. In the following theoretical calculation, we compare them with modes with $l=$ 0, 2, and 3.

Based on above analyses, our mode identifications differ from those of Poretti et al. (2009). Poretti et al. (2009) identified the observed frequencies with the FAMIAS method to fit the line profile variations. There is still uncertain on uniqueness of the solution for the multiparameter fitting method of the line profile variations. Our mode identifications are on basis of the property of g-mode rotational splitting. For the $\delta$ Scuti star HD 50844, rotational splittings for modes with $l\ge3$ are very close according to Eq. (1). Distinguishing multiplets of $l=4$ or higher spherical harmonic degree $l$ from those of $l=3$ is very difficult. In spectroscopy, the behaviour of amplitudes and phases across the line profiles supplies information on both spherical harmonic degree $l$ and azimuthal number $m$. For instance phase diagrams of six detected modes of BV Cir clearly show they are prograde modes with high azimuthal number -14 $\leq m \leq$ -12 (Mantegazza et al. 2001).

\section{Stellar Models}
\subsection{Input Physics}
In our work, we use the package "pulse" from version 6596 of the Modules for Experiments in Stellar Astrophysics (MESA) (Paxton et al. 2011, 2013) to compute stellar evolutionary models and to calculate their pulsation frequencies (Christensen-Dalsgaard 2008). Our theoretical models for HD 50844 are constructed from the ZAMS to the post-main sequence stage, fully covering the observed ranges of gravitational acceleration and effective temperature. The OPAL equation-of-state tables (Rogers $\&$ Nayfonov 2002) are used. The OPAL opacity tables from Iglesias $\&$ Rogers (1996) are used in the high temperature region and opacity tables from Ferguson et al. (2005) are used in the low temperature region. The $T - \tau$ relation of Eddington grey atmosphere is adopted in the atmosphere integration. The mixing-length theory (B$\ddot{o}$hm-Vitense 1958) is adopted to treat convection. Effects of convective overshooting and element diffusion are not considered in our calculations.

\subsection{Details of Model Grids}
The calibrated value of $\alpha=1.77$ for the sun is adopted in our stellar evolutionary models. The evolutionary track of a star on the HR diagram is determined by the stellar mass $M$ and the initial chemical composition $(X,Y,Z)$. In our calculations, we set the initial helium fraction $Y=0.275$ as a constant. Then we choose the range of mass fraction of heavy-elements $Z$ from 0.005 to 0.018, in order to cover the observational value of $\rm [Fe/H]=-0.40\pm0.20$ (Poretti et al. 2005).

A grid of stellar models are computed with MESA, $M$ ranging from 1.5 $M_{\odot}$ to 2.2 $M_{\odot}$ with a step of 0.01 $M_{\odot}$, and $Z$ ranging from 0.005 to 0.018 with a step of 0.001. Figure 1 shows the grid of evolutionary tracks with various sets of $M$ and $Z$. The error box corresponds to the effective temperature range of 7300 K $<$ $T_{\rm eff}$ $<$ 7700 K and to the gravitational acceleration range of 3.40 $<$ log$g$ $<$ 3.80. For each stellar model falling in the error box, we calculate its frequencies of pulsation modes with $l$ = 0, 1, 2 and 3, and fit them to those observational frequencies according to
\begin{equation}
\chi^{2}=\frac{1}{N}\sum(|\nu_{\rm obs}^{i}-\nu_{\rm mod}^{i}|^{2}),
\end{equation}
In Eq. (2), $\nu_{\rm obs}^{i}$ is observational frequency, $\nu_{\rm mod}^{i}$ is calculated pulsation frequency, and $N$ is total number of observational frequencies. Based on numerical simulations, most of the uncertainties of the calculated pulsation frequencies are less than 0.03 $\mu$Hz except that a few of them reach up to 0.06 $\mu$Hz.

\section{Analysis of Results}
\subsection{Best-fitting model}
In Section 2, we give our possible mode identifications for the observed frequencies obtained by Balona (2014) based on the rotational splitting law. When doing model fittings, we try to use the calculated frequencies of each model to fit four identified modes, including 2 modes with $l=1$ ($f_{22}$ and $f_{29}$), 1 mode with $l=2$ ($f_{11}$), and the fundamental radial mode ($f_{4}$). Poretti et al. (2009) suggested that $f_{4}$ might be one mode with $l$ = 0 based on the mode identifications with the FAMIAS method. Besides, Poretti et al.(2009) detected $f_{4}$ in the variations of equivalent width and radial velocity and identified $f_{4}$ as the mode with $l=0$. Moreover, the photometric identifications made by Poretti et al. (2009) on basis of the colour information of the multi-colour photometric data show $f_{4}$ as the fundamental radial mode. This is in accordance with the property that $f_{4}$ has the highest amplitude in the variations of both equivalent width and radial velocity. In our work, we use the identification of $f_{4}$ as the fundamental radial mode.

Figure 2 shows the change of $1/\chi^{2}$ as a function of the effective temperature $T_{\rm eff}$ for all considered models. In Fig. 2, each curve corresponds to one evolutionary track. It can be noticed in Fig. 2 that the value of $1/\chi^{2}$ is very large in a very small parameter space, i.e., $M=1.80 - 1.81$ $M_{\odot}$ and $Z=0.008 - 0.009$. Their physical parameters are very close. The physical parameters of HD 50844 are obtained based on these models. They are listed in Table 3. We select the theoretical model with the minimum value of $\chi^{2}$ corresponding to $(M=1.81,Z=0.008)$ as our best-fitting model, which is marked with the filled cycle in Fig.2.

The theoretical frequencies of our best-fitting model are listed in Table 4, where $n_{p}$ is the number of radial nodes in the p-mode propagation region, and $n_{g}$ the number of radial nodes in the g-mode propagation region. Particularly, $\beta_{k,l}$ is a parameter measuring the size of rotational splitting for rigid body in the general formula of rotational splitting derived by Christensen-Dalsgaard (2003):
\begin{equation}
\beta_{k,l}=\frac{\int_{0}^{R}(\xi_{r}^{2}+l(l+1)\xi_{h}^{2}-2\xi_{r}\xi_{h}-\xi_{h}^{2})r^{2}\rho dr}
{\int_{0}^{R}(\xi_{r}^{2}+l(l+1)\xi_{h}^{2})r^{2}\rho dr}.
\end{equation}
In Eq.(3), $\xi_{r}$ is the radial displacement, $\xi_{h}$ the horizontal displacement, and $\rho$ the local density. Therefore, the effect of rotation is basically determined by the value of $\beta_{k,l}$. For high-order g modes, the terms containing $\xi_{r}$ can be neglected, thus
\begin{equation}
\beta_{k,l} \backsimeq 1 - \frac{1}{l(l+1)},
\end{equation}
which is in agreement with Eq. (1).

Figure 3 shows the plot of $\beta_{k,l}$ versus the theoretical frequency $\nu$ for the best-fitting model. It can be seen that most of values of $\beta_{k,l}$ in Fig. 3 agree well with the value of 0.5 for $l=1$ modes, 0.833 for $l=2$ modes, or 0.917 for $l=3$ modes derived from Eq. (1). These results indicate that the corresponding modes have pronounced g-mode characteristics. On the other hand, $\beta_{k,l}$ of several $l=1$, $l=2$ and $l=3$ modes deviate considerably from the values derived from Eq. (1), indicating that they also possess significant p-mode characteristics.

Table 5 lists results of comparisons of frequencies for those modes in Table 2, where $m\neq0$ modes in columns named by $\nu_{\rm mod}$ are derived from $m=0$ modes based on $P_{rot}$ and $\beta_{k,l}$. The filled circles in Fig. 3 denote corresponding $m=0$ modes in Table 5. It can be seen clearly in Fig. 3 that values of $\beta_{k,l}$ for $m=0$ modes corresponding to $f_{22}$, $f_{29}$, $f_{11}$, $f_{1}$, $f_{15}$ and $f_{34}$ agree well with those derived from Eq.(1). In Multiplet 7, 8, 9, 11 and 12, $m=0$ components are not observed. Values of $\beta_{k,l}$ for corresponding $m=0$ components in Multiplet 7, 9, 11, and 12 are also in good agreement with those derived form Eq. (1). The mode corresponding to $f_{25}$ in Multiplet 3 and corresponding $m=0$ component in Multiplet 10 have slightly larger values of $\beta_{k,l}$ than those derived from Eq. (1). It can be noticed in Table 5 that there are two possible identifications for Multiplet 8, i.e., corresponding to modes of $m$ = (-1, +2) derived from 80.189 $\mu$Hz (2, 0, -78, 0), or $m$=(-2, +1) from 84.368 $\mu$Hz (2, 0, -74, 0). The filled squares in Fig. 3 denote these two possible $m = 0$ modes in Multiplet 8. It can be seen in Fig. 3 that the values of $\beta_{k,l}$ for both of the two choices are in good agreement with Eq. (1). These results confirm our approach of using Eq. (1) to search for rotational splitting in Section 2.

Based on the best-fitting model, possible identifications for $f_{13}$, $f_{16}$, $f_{30}$, $f_{31}$, $f_{32}$, and $f_{40}$ are listed in Table 6. It can be noticed in Table 6 that $f_{30}$, $f_{31}$, $f_{32}$, and $f_{40}$ may be identified as four modes with $l=3$. Poretti er al. (2009) identified $f_{30}$ as $(l=4,m=2)$ and $f_{31}$ as $(l=4, m=3)$. Considering uncertainties of spectroscopic observations obtained by (Poretti et al. 2009), the spherical harmonic degree $l$ of our suggestions agree with those of Poretti et al. (2009). For $f_{16}$, there are three possible identifications, i.e., corresponding to modes of $(l,m)$ =$(2,-2)$, $(3,+2)$, or $(3,0)$. Besides, there are two possible identifications for $f_{15}$ based on the analyses in Section 2. If $f_{15}$ and $f_{13}$ are identified as two components of one incomplete septuplet, $(130.081, 134.465)$ derived from 138.850 $\mu$Hz $(3,2,-62,0)$ or $(130.035,134.392)$ derived from 143.104 $\mu$Hz $(3,2,-60,0)$ may be two possibilities. If $f_{15}$ and $f_{18}$ are identified as two components of one incomplete quintuplet, the results of comparisons are listed in Table 5. Poretti et al. (2009) identified $f_{15}$ as a mode with ($l=8$, $m=5$). The spherical harmonic degree $l$ for both of the two cases ($l=3$ or $2$) are lower than the value of Poretti et al. (2009).
\subsection{Discussions}
An important question is why physical parameters of HD 50844 are well limited in a small region based on four identified pulsation modes. We have found a possible reason to explain this result.

First of all, it should be pointed out that the four identified modes consist of two $l=1$ modes ($f_{22}$ and $f_{29}$), one $l=2$ mode ($f_{11}$), and the fundamental radial mode ($f_{4}$). Table 4 shows that most of pulsation modes are mixed modes. Figure 4 shows the propagation diagram for the best-fitting model. According to the parameter settings of MESA (Paxton et al. 2011, 2013), the boundary of helium core is set to the position where the hydrogen fraction $X_{cb}$ = 0.01. The vertical lines in Fig. 4 and Fig. 5 denote the position of the boundary of the helium core. The inner zone is the helium core, the outer zone is the stellar envelope. It can be seen in Fig. 4 that the Brunt-V$\ddot{a}$is$\ddot{a}$l$\ddot{a}$ frequency $N$ has a peak in the helium core, which corresponds to the hydrogen burning shell. Figure 5 shows distributions of the radial displacement for the fundamental radial mode and the three nonradial modes that we have considered. It can be seen clearly in Fig. 5 that the fundamental radial mode propagates mainly in the stellar envelope, and represents the property of the stellar envelope. However, the three nonradial modes propagate like g modes in the helium core while like p modes in the stellar envelope. This fact confirms that they are mixed modes and can therefore represent the property of the helium core. In order to fit those four modes at the same time, both the stellar envelope and the helium core must be matched to the considered star.

The acoustic radius $T_{\rm h}$ is defined as $T_{\rm h}$ = $\int_{0}^{R}c_{s}^{-1}dr$ (Aerts et al. 2010), where $c_{s}$ is the adiabatic sound speed. Therefore, the acoustic radius $T_{\rm h}$ is mainly determined by the distribution of $c_{s}$ inside the star. The sound speed $c_{s}$ is much smaller in the stellar envelope than in the helium core, thus $T_{\rm h}$ can be used to reflect property of the stellar envelope.

According to the asymptotic theory of g modes, there is an equation for the period separation (Unno et al. 1979; Tassoul 1980)
\begin{equation}
\Delta\bar{P}(l) = \frac{\Pi_{0}}{\sqrt{l(l+1)}} = \frac{2\pi^{2}(\int_{r_{1}}^{r_{2}}\frac{N}{r}dr)^{-1}}{\sqrt{l(l+1)}},
\end{equation}
where $r_{1}$ is the inner boundary of the region where gravity waves propagate, $r_{2}$ is the outer boundary, and $N$ is the Brunt-V$\ddot{a}$is$\ddot{a}$l$\ddot{a}$ frequency. In Eq.(5), $\Pi_{0}$=$2\pi^{2}(\int_{r_{1}}^{r_{2}}N/rdr)^{-1}$, which is mainly determined by the distribution of Brunt-V$\ddot{a}$is$\ddot{a}$l$\ddot{a}$ frequency $N$ in the helium core. Therefore, $\Pi_{0}$ can be used to reflect property of the helium core.

Figure 6 shows the distribution of the period spacing $\Pi_{0}$ versus the acoustic radius $T_{\rm h}$ for theoretical models with the same initial metallicity $Z$ but different stellar mass $M$. Figure 7 shows the same plot for theoretical models with the same stellar mass $M$ but different initial metallicity $Z$. The filled circle corresponds to our best-fitting model, while the filled triangles correspond to stellar models having minimum values of $\chi^{2}$ on the corresponding evolutionary tracks, respectively. It can be noticed in Fig. 6 that the acoustic radius $T_{\rm h}$ of the stellar models marked by the filled triangles obviously deviate from the value of our best-fitting model, which indicates that the stellar envelopes of these models can not match the actual structure of the considered star. In contrast, the period spacing $\Pi_{0}$ of the stellar models marked by the filled triangles in Fig. 7 obviously deviate from the value of our best-fitting model, which indicates that the helium cores of these models can not match the actual structure of the considered star. Based on above arguments further more, the size of the helium core of the $\delta$ Scuti star HD 50844 is determined for the first time. The corresponding physical parameters are listed in Table 3.

Gizon $\&$ Solanki (2003) investigated in details the relation between the stellar oscillation amplitude and the inclination angle $i$ of stellar rotation axes. It can be noticed in Table 1 that the $m=$ 0 component $f_{22}$ of Multiplet 1 has an amplitude that is about 9 times smaller than the $m=-1$ components $f_{21}$ and about 7 times smaller than the $m=+1$ components $f_{23}$. Such large differences correspond to an inclination angle $i\approx76 ^{\circ}$ according to the relation given by Gizon $\&$ Solanki (2003). This fact is roughly in agreement with the value of $82^{\circ}$ (Poretti et al. 2009).  The $m=0$ component in Multiplet 4 has the least amplitude, and the $m=$ 0 component in Multiplet 2 also has a smaller amplitude.

The rotational period $P_{rot}$ is determined to be 2.44$^{+0.13}_{-0.08}$ days according to Eq. (1). It can be noticed in Table 3 that the theoretical radius of HD 50844 is $R$ = 3.300 $\pm$ 0.023 $R_{\odot}$. Then the rotational velocity at the equator is derived as $\upsilon_{\rm rot}=$ 68.33$^{+2.34}_{-3.70}$ km $\rm s^{-1}$ according to $\upsilon_{\rm rot}$ = 2$\pi R/P_{\rm rot}$. Assuming the inclination angle $i=$ 82 $\pm$ 4 deg (Poretti et al. 2009), $\upsilon_{\rm rot}\sin i$ is estimated to be 66.86 $\pm$ 3.64 km $\rm s^{-1}$, which is higher than the value of $\upsilon\sin i$ = 58 $\pm$ 2 km $\rm s^{-1}$ (Poretti et al. 2009). It has been discussed in Sect.4.1 that most of the considered frequencies are mixed modes. They have pronounced g-mode characteristics. The corresponding rotational velocity derived from rotational splitting of these modes mainly reflects rotational properties of the helium core. The $\delta$ Scuti star HD 50844 is a slightly evolved star. As the star evolves into the post-main sequence stage, the core shrinks and the envelope expands. Based on conservation of angular momentum, rotational angular velocity of the core should be larger than that of the envelope. The spectroscopic value of $\upsilon\sin i$ (Poretti et al. 2009) mainly reflects the property of the envelope. This may be the reason why our rotational velocity is higher than that of Poretti et al. (2009).

\section{Summary}
In our work, we have analysed the observed frequencies given by Balona (2014) for possible rotational splitting, and carried out numerical model fittings for the $\delta$ Scuti star HD 50844. We summarize our results as follows:

1. We identify two complete triplets ($f_{21}$, $f_{22}$, $f_{23}$) and ($f_{27}$, $f_{29}$, $f_{33}$) as modes with $l=1$, and one incomplete quintuplet ($f_{9}$, $f_{11}$, $f_{14}$) as modes with $l=2$, as well as one more incomplete triplet ($f_{24}$, $f_{25}$) as modes with $l=1$ and six more incomplete quintuplets ($f_{1}$, $f_{5}$), ($f_{15}$, $f_{18}$), ($f_{35}$, $f_{36}$), ($f_{2}$, $f_{7}$), ($f_{12}$, $f_{20}$), and ($f_{38}$, $f_{39}$) as modes with $l=2$. Besides, three incomplete septuplets $(f_{3}, f_{6}, f_{8})$, $(f_{10}, f_{17}, f_{19})$, and $(f_{26}, f_{28}, f_{34}, f_{37})$ are identified as modes with $l=3$. Based on frequency differences of above multiplets, the corresponding rotational period of HD 50844 is found to be 2.44$^{+0.13}_{-0.08}$ days.

2. Based on our model calculations, we compare theoretical pulsation modes with four identified observational modes, including three nonradial modes ($f_{11}$, $f_{22}$, $f_{29}$) and the fundamental radial mode ($f_{4}$). The physical parameters of HD 50844 are well limited in a small region. Based on the fitting results, the theoretical model with $M=1.81$ $M_{\odot}$, $Z=$ 0.008 is suggested as the best-fitting model.

3. Based on our best-fitting model, we find that values of $\beta_{k,l}$ for most of the calculated modes are in good agreement with the asymptotic values for g modes. Some modes may have values of $\beta_{k,l}$ that are considerably higher than the asymptotic values. However, the values of $\beta_{k,l}$ for the $m=0$ modes in those identified triplets, quintuplets, or septuplets are in good agreement with the asymptotic values of g modes, which confirms that our approach of searching for rotational splitting based on the rule of g modes is self-consistent.

4. Based on comparisons of all observed frequencies with their theoretical counterparts, we find that most of the considered frequencies may belong to mixed modes. The radial fundamental mode $f_{4}$ reflects the property of the stellar envelope, while the three nonradial modes $f_{11}$, $f_{22}$, and $f_{29}$ reflect property of the helium core. These features require that both the stellar envelope and the helium core must be matched to the actual structure in order to fit those four oscillation modes. Finally, the mass of helium core of HD 50844 is estimated to be 0.173 $\pm$ 0.004 $M_{\odot}$.

\begin{acknowledgements}
We are sincerely grateful to an anonymous referee for instructive advice and productive suggestions, which greatly help us to improve the manuscript. This work is funded by the NFSC of China (Grant No. 11333006, 11521303, and 11403094) and by the foundation of Chinese Academy of Sciences (Grant No. XDB09010202 and "Light of West China" Program). We gratefully acknowledge the computing time granted by the Yunnan Observatories, and provided on the facilities at the Yunnan Observatories Supercomputing Platform. We are also very grateful to J.-J. Guo, G.-F. Lin, Q.-S. Zhang, Y.-H. Chen, and J. Su for their kind discussions and suggestions.
\end{acknowledgements}

  \begin{figure}
   \centering
   \includegraphics[width=8cm]{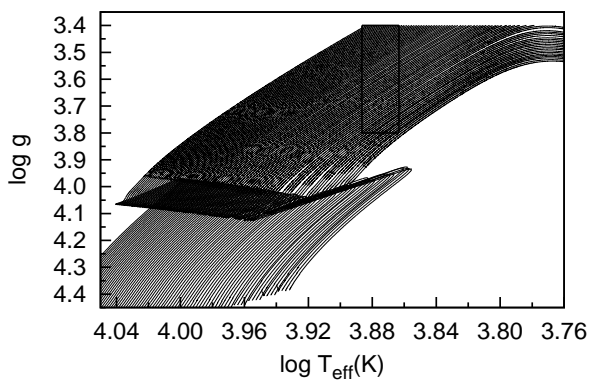}
      \caption{Evolutionary tracks. The rectangle is the 1$\sigma$ error box for the observational constraints.}
         \label{Fig.1}
   \end{figure}

  \begin{figure}
  \centering
  \includegraphics[width=8cm]{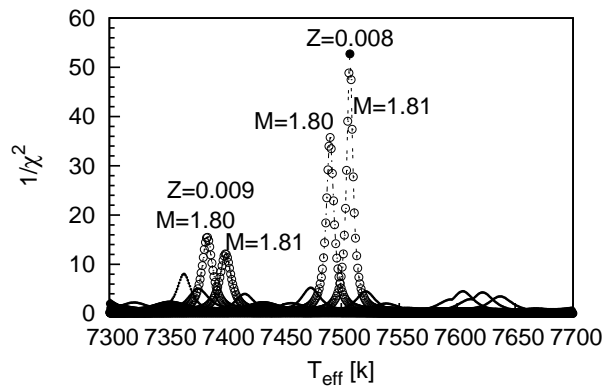}
  \caption{$1/\chi^{2}$ as a function of the effective temperature $T_{\rm eff}$. The filled circle denotes the best-fitting model in Section 4.1.}
  \label{Fig.2}
  \end{figure}

  \begin{figure}
  \centering
  \includegraphics[width=8cm]{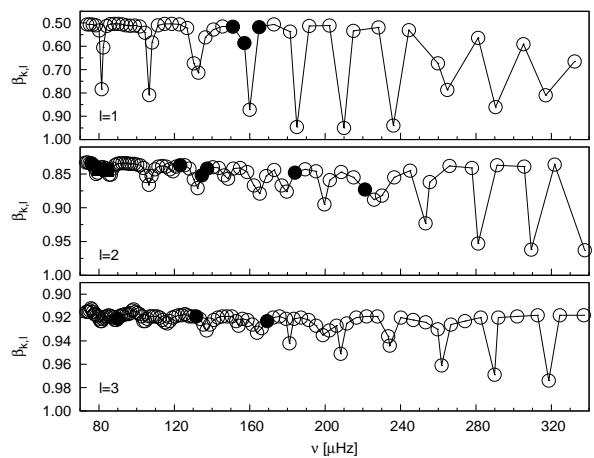}
  \caption{Plot of $\beta_{k,l}$ to theoretical frequency $\nu$ of our best-fitting model. The filled circles denote modes corresponding to the $m=0$ mode in Table 5, and the two filled squares denote modes corresponding to the $m=0$ modes for Multiplet 8.}
  \label{Fig.3}
  \end{figure}

  \begin{figure}
  \centering
  \includegraphics[width=8cm]{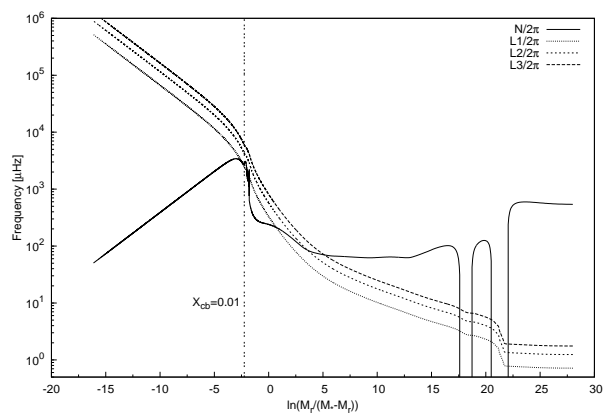}
  \caption{$N$ denotes Brunt$-$V$\ddot{a}$is$\ddot{a}$l$\ddot{a}$ frequency and $L_{l}$ ($l= 1, 2, 3$) denote Lamb frequency. $M_{*}$ denotes the total mass of the star. Vertical line denotes the boundary of the helium core.}
  \label{Fig.4}
  \end{figure}

  \begin{figure}
  \centering
  \includegraphics[width=8cm]{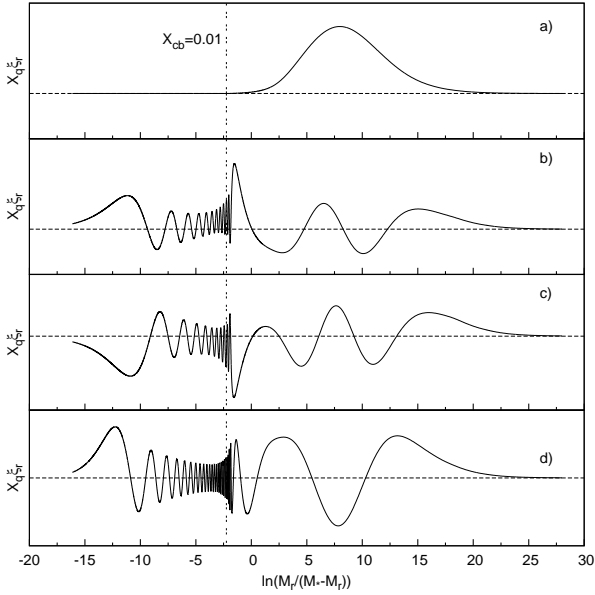}
  \caption{Scaled eigenfunctions of modes corresponding to $f_{4}$, $f_{11}$, $f_{22}$ and $f_{29}$. $X_{q} = \sqrt{q(1-q)}$ and $q=M_{r}/M_{*}$. Panel (a) is for the fundamental radial mode 79.937 $\mu$Hz $(l=0,n_{p}=0,n_{g}=0)$. Panel (b) is for the mode 151.060 $\mu$Hz $(l=1,n_{p}=3,n_{g}=-23)$. Panel (c) is for the mode 164.999 $\mu$Hz $(l=1,n_{p}=4,n_{g}=-21)$. Panel (d) is for the mode 123.113 $\mu$Hz $(l=2,n_{p}=2,n_{g}=-50)$. Vertical line denotes the boundary of the helium core.}
  \label{Fig.5}
  \end{figure}

  \begin{figure}
  \centering
  \includegraphics[width=8cm]{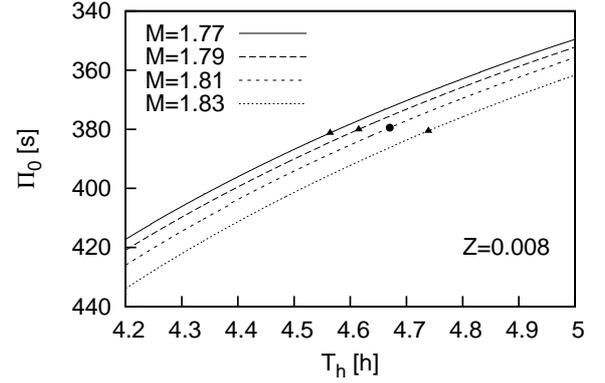}
  \caption{Plot of sound radius $T_{\rm h}$ to period spacing $\Pi_{0}$ with the same initial metallicity $Z=$ 0.008 and different stellar mass $M$. The filled circle represents our best-fitting model. The filled triangles represent models which have minimum value of $\chi^{2}$ on corresponding evolutionary tracks.}
  \label{Fig.6}
  \end{figure}

  \begin{figure}
  \centering
  \includegraphics[width=8cm]{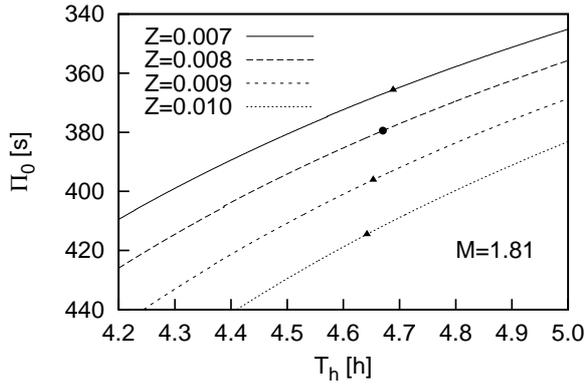}
  \caption{Plot of sound radius $T_{\rm h}$ to period spacing $\Pi_{0}$ with the same stellar mass $M=1.81$ and different initial metallicity $Z$. The filled circle represents our best-fitting model. The filled triangles represent models which have minimum value of $\chi^{2}$ on corresponding evolutionary tracks.}
  \label{Fig.7}
  \end{figure}

  \begin{table*}
  \caption{\label{t1}The independent frequencies of HD 50844 given by Balona (2014). The columns named by ID show the serial number of observed frequencies. Freq. denotes the observed frequency in unit of $\mu$Hz. Ampl. denotes the amplitude in unit of mmag.}
  \centering
  \begin{tabular}{lcccccccc}
  \hline\hline
  ID        &Freq.       &Ampl.    &ID          &Freq.       &Ampl.  \\
            &($\mu$Hz)   &(mmag)   &            &($\mu$Hz)   &(mmag) \\
  \hline
  $f_{1}$   &75.886      &0.82     &$f_{21}$    &148.709     &3.55   \\
  $f_{2}$   &76.141      &0.71     &$f_{22}$    &151.178     &0.40   \\
  $f_{3}$   &76.719      &0.35     &$f_{23}$    &153.627     &2.87   \\
  $f_{4}$   &80.154      &16.73    &$f_{24}$    &154.615     &2.51   \\
  $f_{5}$   &83.958      &0.44     &$f_{25}$    &157.046     &0.85   \\
  $f_{6}$   &85.704      &0.85     &$f_{26}$    &160.300     &0.59   \\
  $f_{7}$   &88.369      &0.66     &$f_{27}$    &162.752     &0.42   \\
  $f_{8}$   &99.116      &0.31     &$f_{28}$    &164.751     &0.56   \\
  $f_{9}$   &115.193     &1.02     &$f_{29}$    &165.088     &0.46   \\
  $f_{10}$  &118.754     &0.79     &$f_{30}$    &167.039     &1.08   \\
  $f_{11}$  &123.030     &0.43     &$f_{31}$    &167.208     &3.08   \\
  $f_{12}$  &129.823     &6.76     &$f_{32}$    &167.375     &0.68   \\
  $f_{13}$  &130.301     &4.37     &$f_{33}$    &167.570     &1.24   \\
  $f_{14}$  &131.227     &0.48     &$f_{34}$    &168.989     &0.46   \\
  $f_{15}$  &134.761     &0.36     &$f_{35}$    &176.206     &0.39   \\
  $f_{16}$  &135.525     &0.69     &$f_{36}$    &180.189     &0.33   \\
  $f_{17}$  &136.007     &1.58     &$f_{37}$    &182.320     &0.42   \\
  $f_{18}$  &138.700     &0.94     &$f_{38}$    &216.702     &0.45   \\
  $f_{19}$  &140.309     &0.70     &$f_{39}$    &228.592     &0.39   \\
  $f_{20}$  &141.648     &3.53     &$f_{40}$    &315.920     &0.66   \\
  \hline
  \end{tabular}
  \end{table*}

\begin{table*}
\footnotesize
\caption{\label{t2}Possible rotational splittings. $\delta\nu$ = frequency spacing in $\mu$Hz}
\centering
\begin{tabular}{lccccccccccc}
\hline\hline
Multiplet &ID      &Freq.     &$\delta\nu$ &$l$  &$m$     &Multiplet &ID      &Freq.     &$\delta\nu$ &$l$ &$m$\\
        &        &($\mu$Hz) &($\mu$Hz)   &     &        &          &        &($\mu$Hz) &($\mu$Hz)   &    &   \\
\hline
        &$f_{21}$&148.709   &            &1    &$-1$    &          &$f_{2}$ &76.141    &            &2   &$(-2,-1)$\\
        &        &          &2.469       &     &        &8         &        &          &12.228\\
1       &$f_{22}$&151.178   &            &1    &$0$     &          &$f_{7}$ &88.369    &            &2   &$(+1,+2)$\\
        &        &          &2.449       &     &        &          &\\
        &$f_{23}$&153.627   &            &1    &$+1$    &          &\\
        &        &          &            &     &        &          &$f_{12}$&129.823   &            &2   &$(-2,-1)$\\
        &        &          &            &     &        &9         &        &          &11.825      &    &\\
        &$f_{27}$&162.752   &            &1    &$-1$    &          &$f_{20}$&141.648   &            &2   &$(+1,+2)$\\
        &        &          &2.336       &     &        &          &\\
2       &$f_{29}$&165.088   &            &1    &$0$     &          &\\
        &        &          &2.482       &     &        &          &$f_{38}$&216.702   &            &2   &$(-2,-1)$\\
        &$f_{33}$&167.570   &            &1    &$+1$    &10        &        &          &11.890      &    &\\
        &        &          &            &     &        &          &$f_{39}$&228.592   &            &2   &$(+1,+2)$\\
        &        &          &            &     &        &          &\\
        &$f_{24}$&154.615   &            &1    &($0$ or $-1$)&     &\\
3       &        &          &2.431       &     &             &     &$f_{3}$ &76.719    &            &3   &$(-3,-2)$\\
        &$f_{25}$&157.046   &            &1    &($+1$ or $0$)&     &        &          &8.985       &    &\\
        &        &          &            &     &             &11   &$f_{6}$ &85.704    &            &3   &$(-1,0)$\\
        &        &          &            &     &             &     &        &          &13.412      &    &\\
        &$f_{9}$ &115.193   &            &2    &$-2$         &     &$f_{8}$ &99.116    &            &3   &$(+2,+3)$\\
        &        &          &7.837       &     &             &\\
4       &$f_{11}$&123.030   &            &2    &0            &\\
        &        &          &8.197       &     &             &     &$f_{10}$&118.754   &            &3   &$(-3,-2)$\\
        &$f_{14}$&131.227   &            &2    &$+2$         &     &        &          &17.253      &    &\\
        &        &          &            &     &             &12   &$f_{17}$&136.007   &            &3   &$(+1,+2)$ \\
        &        &          &            &     &             &     &        &          &4.302       &    &\\
        &$f_{1}$ &75.886    &            &2    &$(-2,-1,0)$  &     &$f_{19}$&140.309   &            &3   &$(+2,+3)$\\
5       &        &          &8.072       &     &             &\\
        &$f_{5}$ &83.958    &            &2    &$(0,+1,+2)$  &\\
        &        &          &            &     &             &     &$f_{26}$&160.300   &            &3   &$(-3,-2)$\\
        &        &          &            &     &             &     &        &          &4.451       &    &\\
        &$f_{15}$&134.761   &            &2   &$(-2,-1,0,+1)$&     &$f_{28}$&164.751   &            &3   &$(-2,-1)$\\
6       &        &          &3.939       &     &             &13   &        &          &4.238       &    &\\
        &$f_{18}$&138.700   &            &2   &$(-1,0,+1,+2)$&     &$f_{34}$&168.989   &            &3   &$(-1,0)$\\
        &        &          &            &     &             &     &        &          &13.331      &    &\\
        &        &          &            &     &             &     &$f_{37}$&182.320   &            &3   &$(+2,+3)$\\
        &$f_{35}$&176.206   &            &2    &$(-2,-1,0,+1)$      \\
7       &        &          &3.983       &     &         \\
        &$f_{36}$&180.189   &            &2    &$(-1,0,+1,+2)$      \\
\hline
\end{tabular}
\end{table*}

\begin{table*}
\caption{\label{t3} The fundamental parameters of the $\delta$ Scuti star HD 50844.}
\centering
\begin{tabular}{lcc}
\hline\hline
Parameter             &Values \\
\hline
$T_{\rm eff} [K]$     &7508  $\pm$ 125 \\
log$g$                &3.658 $\pm$ 0.004\\
$M/M_{\odot}$         &1.81  $\pm$ 0.01\\
$Z$                   &0.008 $\pm$ 0.001\\
$R/R_{\odot}$         &3.300 $\pm$ 0.023\\
$L/L_{\odot}$         &30.98 $\pm$ 2.39\\
$M_{He}/M_{\odot}$    &0.173 $\pm$ 0.004\\
$R_{He}/R_{\odot}$    &0.068 $\pm$ 0.001\\
\hline
\end{tabular}
\end{table*}

\begin{table*}
\footnotesize
\caption{\label{t4}The theoretical frequencies of the best-fitting model. $\nu_{\rm mod}$ represents model frequency in $\mu$Hz. $n_{p}$ is the number of radial nodes in propagation cavity of p-mode. $n_{g}$ is the number of radial nodes in propagation cavity of g-mode. $\beta_{k,l}$ is one parameter weight the size of rotational splitting.}
\centering
\begin{tabular}{lcccccccccc}
\hline\hline
$\nu_{\rm mod}(l,n_{p},n_{g})$ &$\beta_{k,l}$  &$\nu_{\rm mod}(l,n_{p},n_{g})$ &$\beta_{k,l}$  &$\nu_{\rm mod}(l,n_{p},n_{g})$  &$\beta_{k,l}$ &$\nu_{\rm mod}(l,n_{p},n_{g})$ &$\beta_{k,l}$\\
($\mu$Hz)          &     &($\mu$Hz)          &     &($\mu$Hz)          &     &($\mu$Hz)          &\\
\hline
  79.937(0, 0,   0)&     &  74.128(2, 0, -85)&0.833& 221.105(2, 6, -27)&0.873& 116.235(3, 1, -75)&0.925\\
 103.356(0, 1,   0)&     &  75.014(2, 0, -84)&0.834& 225.913(2, 6, -26)&0.888& 117.557(3, 2, -75)&0.923\\
 128.796(0, 2,   0)&     &  75.922(2, 0, -83)&0.834& 230.117(2, 6, -25)&0.882& 119.007(3, 2, -74)&0.920\\
 154.668(0, 3,   0)&     &  76.841(2, 0, -82)&0.836& 236.592(2, 6, -24)&0.855& 120.572(3, 2, -73)&0.919\\
 179.930(0, 4,   0)&     &  77.728(2, 0, -81)&0.841& 245.080(2, 6, -23)&0.845& 122.221(3, 2, -72)&0.918\\
 204.942(0, 5,   0)&     &  78.510(2, 0, -80)&0.850& 253.222(2, 7, -23)&0.923& 123.923(3, 2, -71)&0.918\\
 230.740(0, 6,   0)&     &  79.277(2, 0, -79)&0.848& 255.491(2, 7, -22)&0.862& 125.593(3, 2, -70)&0.917\\
 257.460(0, 7,   0)&     &  80.189(2, 0, -78)&0.842& 265.949(2, 7, -21)&0.838& 126.824(3, 2, -69)&0.918\\
 284.964(0, 8,   0)&     &  81.191(2, 0, -77)&0.840& 277.822(2, 8, -21)&0.841& 127.976(3, 2, -68)&0.919\\
 313.007(0, 9,   0)&     &  82.234(2, 0, -76)&0.840& 281.162(2, 8, -20)&0.953& 129.714(3, 2, -67)&0.919\\
 341.328(0,10,   0)&     &  83.301(2, 0, -75)&0.841& 291.148(2, 8, -19)&0.837& 131.609(3, 2, -66)&0.919\\
                   &     &  84.368(2, 0, -74)&0.844& 305.609(2, 9, -19)&0.839& 133.536(3, 2, -65)&0.921\\
  73.770(1, 0, -49)&0.506&  85.377(2, 0, -73)&0.851& 309.331(2, 9, -18)&0.962& 135.400(3, 2, -64)&0.926\\
  75.163(1, 0, -48)&0.506&  86.294(2, 1, -73)&0.851& 321.765(2, 9, -17)&0.836& 137.101(3, 2, -63)&0.931\\
  76.709(1, 0, -47)&0.507&  87.298(2, 1, -72)&0.842& 337.707(2,10, -17)&0.963& 138.850(3, 2, -62)&0.926\\
  78.364(1, 0, -46)&0.510&  88.458(2, 1, -71)&0.837&                   &     & 140.875(3, 2, -61)&0.922\\
  80.079(1, 0, -45)&0.532&  89.711(2, 1, -70)&0.835&  73.365(3, 0,-122)&0.915& 143.104(3, 2, -60)&0.920\\
  81.427(1, 0, -44)&0.784&  91.017(2, 1, -69)&0.834&  73.966(3, 0,-121)&0.915& 145.460(3, 2, -59)&0.919\\
  82.262(1, 0, -43)&0.605&  92.357(2, 1, -68)&0.834&  74.573(3, 0,-120)&0.915& 147.905(3, 3, -59)&0.919\\
  84.045(1, 1, -43)&0.513&  93.700(2, 1, -67)&0.834&  75.176(3, 0,-119)&0.914& 150.404(3, 3, -58)&0.919\\
  86.042(1, 1, -42)&0.505&  94.961(2, 1, -66)&0.834&  75.741(3, 0,-118)&0.912& 152.752(3, 3, -57)&0.923\\
  88.135(1, 1, -41)&0.504&  96.060(2, 1, -65)&0.835&  76.259(3, 0,-117)&0.913& 154.085(3, 3, -56)&0.927\\
  90.289(1, 1, -40)&0.504&  97.268(2, 1, -64)&0.835&  76.826(3, 0,-116)&0.915& 156.107(3, 3, -55)&0.921\\
  92.440(1, 1, -39)&0.506&  98.704(2, 1, -63)&0.835&  77.463(3, 0,-115)&0.917& 158.756(3, 3, -54)&0.922\\
  94.501(1, 1, -38)&0.510& 100.261(2, 1, -62)&0.836&  78.126(3, 0,-114)&0.917& 161.419(3, 3, -53)&0.926\\
  96.588(1, 1, -37)&0.512& 101.882(2, 1, -61)&0.837&  78.795(3, 0,-113)&0.918& 163.877(3, 3, -52)&0.933\\
  98.956(1, 1, -36)&0.511& 103.525(2, 1, -60)&0.841&  79.467(3, 0,-112)&0.919& 166.324(3, 3, -51)&0.929\\
 101.614(1, 1, -35)&0.515& 105.094(2, 1, -59)&0.853&  80.136(3, 0,-111)&0.921& 169.171(3, 3, -50)&0.923\\
 104.448(1, 1, -34)&0.542& 106.458(2, 1, -58)&0.866&  80.781(3, 0,-110)&0.923& 172.358(3, 3, -49)&0.920\\
 106.628(1, 1, -33)&0.809& 107.890(2, 1, -57)&0.853&  81.405(3, 0,-109)&0.923& 175.756(3, 4, -49)&0.919\\
 108.119(1, 2, -33)&0.584& 109.624(2, 1, -56)&0.843&  82.070(3, 0,-108)&0.921& 179.227(3, 4, -48)&0.921\\
 111.257(1, 2, -32)&0.510& 111.535(2, 1, -55)&0.839&  82.795(3, 0,-107)&0.920& 181.057(3, 4, -47)&0.942\\
 114.802(1, 2, -31)&0.504& 113.545(2, 2, -55)&0.838&  83.560(3, 0,-106)&0.919& 183.359(3, 4, -46)&0.921\\
 118.595(1, 2, -30)&0.504& 115.605(2, 2, -54)&0.839&  84.349(3, 0,-105)&0.919& 187.183(3, 4, -45)&0.920\\
 122.602(1, 2, -29)&0.506& 117.600(2, 2, -53)&0.843&  85.157(3, 0,-104)&0.918& 191.181(3, 4, -44)&0.922\\
 126.726(1, 2, -28)&0.522& 119.260(2, 2, -52)&0.846&  85.981(3, 0,-103)&0.919& 195.152(3, 4, -43)&0.927\\
 130.331(1, 2, -27)&0.673& 120.950(2, 2, -51)&0.839&  86.813(3, 0,-102)&0.919& 198.767(3, 4, -42)&0.935\\
 132.705(1, 2, -26)&0.713& 123.113(2, 2, -50)&0.837&  87.637(3, 0,-101)&0.920& 202.195(3, 5, -42)&0.931\\
 136.429(1, 3, -26)&0.562& 125.507(2, 2, -49)&0.837&  88.426(3, 0,-100)&0.922& 206.064(3, 5, -41)&0.927\\
 140.721(1, 3, -25)&0.528& 127.980(2, 2, -48)&0.842&  89.186(3, 1,-100)&0.922& 208.194(3, 5, -40)&0.951\\
 145.499(1, 3, -24)&0.515& 130.328(2, 2, -47)&0.858&  89.977(3, 1, -99)&0.921& 211.473(3, 5, -39)&0.925\\
 151.060(1, 3, -23)&0.516& 132.312(2, 2, -46)&0.871&  90.837(3, 1, -98)&0.919& 216.489(3, 5, -38)&0.920\\
 157.146(1, 3, -22)&0.586& 134.571(2, 2, -45)&0.852&  91.753(3, 1, -97)&0.918& 221.926(3, 5, -37)&0.919\\
 159.957(1, 3, -21)&0.872& 137.350(2, 2, -44)&0.842&  92.707(3, 1, -96)&0.918& 227.679(3, 6, -37)&0.919\\
 164.999(1, 4, -21)&0.518& 140.383(2, 3, -44)&0.840&  93.688(3, 1, -95)&0.917& 233.424(3, 6, -36)&0.936\\
 172.796(1, 4, -20)&0.506& 143.525(2, 3, -43)&0.841&  94.688(3, 1, -94)&0.917& 234.266(3, 6, -35)&0.944\\
 181.381(1, 4, -19)&0.537& 146.442(2, 3, -42)&0.852&  95.696(3, 1, -93)&0.917& 240.137(3, 6, -34)&0.920\\
 184.971(1, 4, -18)&0.946& 148.490(2, 3, -41)&0.857&  96.678(3, 1, -92)&0.916& 246.672(3, 6, -33)&0.922\\
 191.532(1, 5, -18)&0.513& 151.285(2, 3, -40)&0.842&  97.559(3, 1, -91)&0.914& 253.285(3, 6, -32)&0.924\\
 202.262(1, 5, -17)&0.511& 154.841(2, 3, -39)&0.841&  98.379(3, 1, -90)&0.913& 259.690(3, 7, -32)&0.930\\
 209.971(1, 5, -16)&0.950& 158.600(2, 3, -38)&0.847&  99.335(3, 1, -89)&0.915& 261.738(3, 7, -31)&0.961\\
 214.936(1, 6, -16)&0.534& 162.205(2, 3, -37)&0.867& 100.404(3, 1, -88)&0.916& 266.698(3, 7, -30)&0.926\\
 228.286(1, 6, -15)&0.519& 165.256(2, 3, -36)&0.879& 101.508(3, 1, -87)&0.917& 274.132(3, 7, -29)&0.923\\
 236.146(1, 6, -14)&0.940& 168.714(2, 4, -36)&0.853& 102.583(3, 1, -86)&0.919& 282.527(3, 7, -28)&0.920\\
 244.380(1, 7, -14)&0.531& 172.920(2, 4, -35)&0.844& 103.550(3, 1, -85)&0.923& 289.899(3, 8, -28)&0.969\\
 259.795(1, 7, -13)&0.673& 176.887(2, 4, -34)&0.867& 104.500(3, 1, -84)&0.923& 291.819(3, 8, -27)&0.920\\
 264.796(1, 7, -12)&0.787& 179.615(2, 4, -33)&0.876& 105.609(3, 1, -83)&0.921& 301.852(3, 8, -26)&0.919\\
 281.150(1, 8, -12)&0.564& 183.967(2, 4, -32)&0.848& 106.840(3, 1, -82)&0.919& 312.725(3, 9, -26)&0.918\\
 290.383(1, 8, -11)&0.861& 189.361(2, 4, -31)&0.843& 108.142(3, 1, -81)&0.919& 318.626(3, 9, -25)&0.974\\
 305.164(1, 9, -11)&0.591& 195.099(2, 5, -31)&0.846& 109.493(3, 1, -80)&0.919& 324.479(3, 9, -24)&0.918\\
 316.966(1, 9, -10)&0.810& 199.571(2, 5, -30)&0.895& 110.875(3, 1, -79)&0.919& 337.177(3, 9, -23)&0.918\\
 332.335(1,10, -10)&0.665& 202.435(2, 5, -29)&0.859& 112.270(3, 1, -78)&0.920\\
                   &     & 208.527(2, 5, -28)&0.847& 113.646(3, 1, -77)&0.922\\
  73.274(2, 0, -86)&0.833& 215.073(2, 5, -27)&0.855& 114.966(3, 1, -76)&0.924\\
\hline
\end{tabular}
\end{table*}

\begin{table*}
\footnotesize
\caption{\label{t5}Result of comparisons of frequencies for those modes in Table 2. $\nu_{\rm obs}$ shows observed frequencies in $\mu$Hz, $\nu_{\rm mod}$ shows theoretical frequencies in $\mu$Hz. $\Delta\nu$ = $|\nu_{\rm obs}-\nu_{\rm mod}|$}
\centering
\begin{tabular}{lccccccccccc}
\hline\hline
Multiplet &ID  &$\nu_{obs}$&$\nu_{\rm mode}$  &$\Delta\nu$ &Multiplet &ID &$\nu_{obs}$  &$\nu_{\rm mode}$  &$\Delta\nu$\\
          &    &($\mu$Hz)  &($\mu$Hz)         &($\mu$Hz)   &          &   &($\mu$Hz)    &($\mu$Hz)         &($\mu$Hz)\\
\hline
         &$f_{21}$&148.709   &148.617(1,-1)   &0.092       &          &$f_{2}$ &76.141    &76.202(2,-1)    &0.061\\
         &        &          &                &            &$8^{a}$   &        &          &                &\\
1        &$f_{22}$&151.178   &151.060(1,0)    &0.118       &          &$f_{7}$ &88.369    &88.163(2,+2)    &0.206\\
         &        &          &                &            &\\
         &$f_{23}$&153.627   &153.503(1,+1)   &0.124       &\\
         &        &          &                &            &          &$f_{2}$ &76.141    &76.375(2,-2)    &0.234\\
         &        &          &                &            &$8^{b}$   &        &          &                &\\
         &$f_{27}$&162.752   &162.546(1,-1)   &0.206       &          &$f_{7}$ &88.369    &88.364(2,+1)    &0.005\\
         &        &          &                &            &\\
2        &$f_{29}$&165.088   &164.999(1,0)    &0.089       &\\
         &        &          &                &            &          &$f_{12}$&129.823   &129.376(2,-2)   &0.447\\
         &$f_{33}$&167.570   &167.452(1,+1)   &0.118       &9         &        &          &                &\\
         &        &          &                &            &          &$f_{20}$&141.648   &141.337(2,+1)   &0.311\\
         &        &          &                &            &\\
         &$f_{24}$&154.615   &154.371(1,-1)   &0.244       &\\
3        &        &          &                &            &          &$f_{38}$&216.702   &216.971(2,-1)   &0.269\\
         &$f_{25}$&157.046   &157.146(1,0)    &0.100       &10        &        &          &                &\\
         &        &          &                &            &          &$f_{39}$&228.592   &229.372(2,+2)   &0.780\\
         &        &          &                &            &\\
         &$f_{9}$ &115.193   &115.187(2,-2)   &0.006       &\\
         &        &          &                &            &          &$f_{3}$ &76.719    &76.894(3,-3)    &0.175\\
4        &$f_{11}$&123.030   &123.113(2,0)    &0.083       &          &        &          &                &\\
         &        &          &                &            &11        &$f_{6}$ &85.704    &85.616(3,-1)    &0.088\\
         &$f_{14}$&131.227   &131.039(2,+2)   &0.188       &          &        &          &                &\\
         &        &          &                &            &          &$f_{8}$ &99.116    &98.699(3,+2)    &0.417\\
         &        &          &                &            &\\
         &$f_{1}$ &75.886    &75.922(2,0)     &0.036       &\\
5        &        &          &                &            &          &$f_{10}$&118.754   &118.555(3,-3)   &0.199\\
         &$f_{5}$ &83.958    &83.820(2,+2)    &0.138       &          &        &          &                &\\
         &        &          &                &            &12        &$f_{17}$&136.007   &135.960(3,+1)   &0.047\\
         &        &          &                &            &          &        &          &                &\\
         &$f_{15}$&134.761   &134.571(2,0)    &0.190       &          &$f_{19}$&140.309   &140.312(3,+2)   &0.003\\
6        &        &          &                &            &\\
         &$f_{18}$&138.700   &138.605(2,+2)   &0.095       &\\
         &        &          &                &            &          &$f_{26}$&160.300   &160.430(3,-2)   &0.130\\
         &        &          &                &            &          &        &          &                &\\
         &$f_{35}$&176.206   &175.936(2,-2)   &0.270       &          &$f_{28}$&164.751   &164.801(3,-1)   &0.050\\
7        &        &          &                &            &13        &        &          &                &\\
         &$f_{36}$&180.189   &179.952(2,-1)   &0.237       &          &$f_{34}$&168.989   &169.171(3,0)    &0.182\\
         &        &          &                &            &          &        &          &                &\\
         &        &          &                &            &          &$f_{37}$&182.320   &182.282(3,+3)   &0.038\\
\hline
\end{tabular}
\end{table*}

\begin{table*}
\caption{\label{t6}Possible mode identifications for the rest of observed frequencies based on our best-fitting model. $\Delta\nu$ = $|\nu_{\rm obs}-\nu_{\rm mod}|$.}
\centering
\begin{tabular}{llll}
\hline\hline
ID &$\nu_{\rm {obs}}$&$\nu_{\rm {mod}}(l,n_{\rm p},n_{\rm g},m)$ &$\Delta\nu$\\
        &($\mu$Hz)&($\mu$Hz)           &($\mu$Hz)\\
\hline
$f_{4}$ &80.154   &79.937(0,0,0)       &0.217\\
        &\\
$f_{13}$&130.301  &130.331(1,2,-15,0)  &0.030\\
        &         &130.328(2,2,-47,0)  &0.027\\
        &\\
$f_{16}$&135.525  &135.561(2,2,-43,-2) &0.036\\
        &         &135.517(3,2,-69,+2) &0.008\\
        &         &135.400(3,2,-64,0)  &0.125\\
        &\\
$f_{30}$&167.039  &167.053(3,4,-49,-2) &0.014\\
        &\\
$f_{31}$&167.208  &167.253(3,3,-56,+3) &0.045\\
        &\\
$f_{32}$&167.375  &167.487(3,3,-54,+2) &0.112\\
        &\\
$f_{40}$&315.920  &315.786(3,9,-24,-2) &0.134\\
\hline
\end{tabular}
\end{table*}
\end{document}